\documentstyle[12pt,epsf]{article}
\newcommand{\be}{\begin{equation}}
\newcommand{\ee}{\end{equation}}
\newcommand{\bea}{\begin{eqnarray}}
\newcommand{\eea}{\end{eqnarray}}
\newcommand{\dv}{\delta v}
\newcommand{\la}{\left\langle}
\newcommand{\ra}{\right\rangle}
\newcommand{\dmf}{d\mu_{\mbox{\tiny $R$,$L_0$}}(h)}

\begin{document}
\title{Multi-time, multi-scale correlation functions  in turbulence
and in turbulent  models}

\author{L. Biferale$^1$, G. Boffetta$^2$, A. Celani$^3$
and F. Toschi$^{4}$}

\maketitle
\centerline{ $^1$Dipartimento di Fisica, Universit\`a "Tor Vergata",}
\centerline{ Via della Ricerca Scientifica 1, I-00133 Roma, Italy, and 
INFM, Unit\`a di Tor Vergata.}
\centerline{ $^2$Dipartimento di Fisica Generale, Universit\`a di Torino,}
\centerline{ Via Pietro Giuria 1, I-10125 Torino, Italy, and 
INFM, Unit\`a di Torino Universit\`a.}
\centerline{$^3$Dipartimento di Ingegneria Aerospaziale, Politecnico di Torino}
\centerline{Corso Duca degli Abruzzi 24, 10129 Torino, Italy,}
\centerline{and INFM, Unit\`a di Torino.}
\centerline{$^4$Dipartimento di Fisica, Universit\`a di Pisa}
\centerline{Piazza Torricelli 2, I-56126, Pisa, Italy,}
\centerline{and INFM, Unit\`a di Tor Vergata.}

 \begin{abstract}
A multifractal-like
representation for  multi-time multi-scale velocity
correlation in turbulence and dynamical turbulent models
is proposed. 
The importance of subleading contributions to time correlations  
is highlighted. The fulfillment of the
dynamical constraints due to the equations of motion is thoroughly
discussed.
The prediction stemming from this representation are
tested within the framework of shell models for turbulence. 
\end{abstract}
\section{Introduction}
\label{sec:1}

Turbulent flows are characterized by a highly chaotic and intermittent
transfer of fluctuations from the stirring length, {\it outer} scale, $L_0$,
 down to  the  
viscous dissipation length, {\it inner} scale, $l_d$.
 The Reynolds number
defines the ratio between the outer and the inner scales: $L_0/l_d =Re^{3/4}$.
We talk about fully developed turbulent flows in the limit 
$Re \rightarrow \infty$, in this limit it is safely assumed that 
there exists an inertial range of scales, $l_d \ll r \ll L_0$, where
the time evolution  feels only the non-linear terms  
of the Navier-Stokes eqs.\\

The highly chaotic and intermittent transfer of energy 
leads to  non-trivial correlation among fluctuations of the velocity
fields at different scales and at different time-delays \cite{frisch} .\\

The natural set of observable which one would like to  control are 
the following:
\be
C^{p,q}\left(r,R|t\right) = \la\dv_r^p(t) \cdot \dv_R^q(0) \ra
\label{cf0} 
\ee
where $\dv_r(t) = v(x+r,t)-v(x,t)$
and $l_d \ll r < R \ll L_0$. In (\ref{cf0})
  we have, for sake of simplicity,
neglected the vectorial and tensorial dependencies in the velocity
fields and velocity  correlations respectively. 

Some subclasses of the Multi-Scale Multi-Time (MSMT)
correlation functions  (\ref{cf0}) have recently attracted the 
attention of many scientists 
\cite{eyink,procaccia,procaccia2,bbt,kadanoff}.
\noindent
By evaluating (\ref{cf0}) with  $r=R$, at changing $R$,
and at  zero-time delay, $t=0$,
 we have the celebrated Structure Functions (SF) of order $q+p$.
Further, we may also investigate  Multi-Scale (MS)
correlation functions when we have different lengths involved $r \neq
R$ at zero delay, $t=0$ as well as  single-scale correlation functions
(CF) by fixing  $r=R$ at varying time delay $t$ etc...

Structure functions have been, so far, the most studied turbulent quantities
(see \cite{frisch} for a recent theoretical and experimental review).
On the other hand, 
only recently some theoretical and experimental efforts have been
done in order to understand the time properties of single scale (CF)
correlations, $C^{p,q}\left(r,r\,|t\right)$ \cite{procaccia,procaccia2} and 
the scaling properties of multi-scale correlations (MS)
at zero time delay, $C^{p,q}\left(r,R\,|0\right)$
 \cite{eyink,procaccia,bbt}. \

In this Paper, we  propose and check  a general phenomenological framework
capable to capture all the above
 mentioned  correlation functions and in agreement
with the typical structure of non-linear terms of Navier-Stokes eqs. 
In Section \ref{sec:2} the framework of the multifractal
description of correlations is briefly recalled and critically examined.
In Section \ref{sec:3} a representation for single scale 
time correlations is introduced, and its  predictions 
are tested within the framework of shell models of turbulence.
In Section \ref{sec:4} we deal with the most generic two-scales time 
correlation.

\section{Background: the multifractal description of time
correlations}
\label{sec:2}
One of the most important  outcomes of experimental and theoretical
analysis of  turbulent flows is the spectacular 
ability of simple multifractal phenomenology \cite{frisch,pf} to capture 
the leading behavior of structure functions and
of multi-scale correlation functions at zero-time delays
\cite{eyink,procaccia,bbt}. This may appear not surprising because,
as far as time-delays are not concerned, one may 
expect that 
(many) different phenomenological descriptions may well reproduce
scaling laws typical of (SF) and of (MS) functions: multifractals
being just one of these descriptions.  More striking were the recent 
findings \cite{procaccia} that multifractal phenomenology may
easily be extended to the time-domain such as to give a
precise prediction on the behavior of the time properties
of single scale correlations. As soon as
time enters in the game, one must ask consistency with the equation
of motion: the major break-through was that one may write a  time-multifractal
description in agreement with the equation of motion\footnote{An
important remark is now in order: when we refer to
time-properties of turbulent flows we always mean the
time-properties of the velocity fields once the trivial sweeping
effects of large scale on small scales is removed, for
example by choosing to work in a quasi-Lagrangian
reference frame \cite{procaccia}.}.
 
Let us now quickly enter in the details of previous 
findings \cite{procaccia} in order to clarify both the phenomenological
framework and  the notation that we will
use in the following.

Let us remind that the multifractal (Parisi-Frisch) description of 
single-time correlation functions is based on the assumptions that 
inertial range statistics is fully determined by a cascade process 
conditioned to some large scale configuration:
\be
\dv_r = W(r,R) \cdot \dv_R
\ee
where the fluctuating  function $W(r,R)$ 
can be expressed in terms of  a 
superposition of local scaling solution $W(r,R) \sim  (\frac{r}{R})^{h(x)}$ 
with a scaling exponent $h(x)$ which assumes different values
$h$ in a class of 
interwoven fractal sets with fractal codimension $Z(h) = 3-D(h)$.
From this assumption one can write the
expression for any 
structure functions of order $m$, which in  our notation ($m=p+q$) 
becomes:
\bea
S^m(R) \equiv C^{p,q}\left(R,R|0\right) \sim \la W\left(R,L_0\right)^m\ra \la U_0^m\ra\\
\equiv \la U_0^m\ra\int \dmf \left(\frac{R}{L_0}\right)^{h\,m}
\sim \left(\frac{R}{L_0}\right)^{\zeta(m)}
\label{eq:pf}
\eea
 where we have introduced the shorthand notation
 $\dmf\equiv dh \left(\frac{R}{L_0}\right)^{Z(h)}$
to define the probability of having a local exponent $h$
connecting fluctuations between scales $R$ and $L_0$.
In (\ref{eq:pf}) we have used a steepest descent estimate, 
in the limit $R/L_0 \rightarrow 0$, in order to define the intermittent 
scaling exponents $\zeta(m)$. Intensity of intermittency depends on the departure of the $\zeta(m)$ exponents from a linear behavior in $m$.\\ 
In order to extend this description to the time domain,
Procaccia and coll. have proposed to consider that 
two velocity fluctuations, both  at scale $R$ but separated by a time
delay $t$,  can be thought to be characterized by the
same  fragmentation process
$W_{R,L_0}(t)\sim W_{R,L_0}(0)$ as long as the time separation
$t$ is smaller then the "instantaneous" eddy-turn-over time of that scale,
 $\tau_R$, 
while they must be almost  uncorrelated for time larger then $\tau_R$.
Considering that the eddy-turn-over time at scale $R$ is itself
a fluctuating quantity $\tau_R \sim R/(\dv_R) \sim R^{1-h}$ we may write
down \cite{procaccia}:
\be
C^{p,q}(R,R|t)
 \sim \int \dmf \left(\frac{R}{L_0}\right)^{h(p+q)}
F_{p,q}\left(\frac{t}{\tau_R}\right)
\label{cf}
\ee
where the time-dependency is hidden in the function $f(x)$
which must be a smooth function of its argument (for example
a decreasing exponential). \\
From (\ref{cf}) is it straightforward to realize that at zero-time
separation we recover the usual SF representation. It is much more interesting
to notice that (\ref{cf}) is also  in agreement with the constraints
imposed by the non-linear part of the Navier-Stokes eqs.
 Indeed, to make short  a long story (see \cite{procaccia2}
for a rigorous discussion) we may say that 
under the only hypothesis that non-linear terms are dominated
by local interactions in the Fourier space we can safely
assume that as far as power-law counting is concerned
the inertial  terms of 
Navier-Stokes equations for the velocity difference $\dv_R$
can be estimated to be of the form:
\begin{equation}
\partial_t \dv_R(t) \sim O\left[\frac{\left(\dv_R(t)\right)^2}{R}\right]
\ee
and therefore we may check that:
\bea
\partial_t C^{p,q}(R,R|t)
\sim \int \dmf  \left(\frac{R}{L_0}\right)^{h(p+q)} \left(\tau_R\right)^{-1} 
F_{p,q}'\left(\frac{t}{\tau_R}\right)
\sim {C^{p+1,q}(R,R|t) \over R}
\label{cft}
\eea
where of course in the last relation there is  hidden the famous
closure-problem of turbulence, now restated in term of the
relation : 
$\frac{d}{dt} F_{p,q}(t) \sim F_{p+1,q}(t)$.
Let us therefore stress that we are ``not solving turbulence''
but just building up a phenomenological framework where all the leading
(and sub-leading, see below) scaling properties are 
consistent with the constraints imposed by the equation of
motion\footnote{In order to really attack
 the NS equations in this framework one
should dive into the structure of the $F_{pq}$-functions in great detail:
a problem  which seems still to be far from convergence \cite{procaccia2}.}.

In the following we shall show how the representation
(\ref{cf}) must be improved to encompass the most general 
 multi-time multi-scale correlation 
$C^{p,q}\left(r,R|t\right)$.

\section{Single scale time correlations}
\label{sec:3}

We shall first show in which respect the expression (\ref{cf})
may not be considered a satisfactory representation of 
single scale time correlation.
The first comment that can be raised about 
(\ref{cf}) is that 
it misses important sub-leading terms which may completely
spoil the long-time scaling behaviour:
indeed, the main hypothesis that correlation $C^{p,q}(R,R|t)$ feels
only the eddy-turn-over time of the scale $R$ is too strong.
It is actually correct only  when the correlation function
has zero disconnected part, i.e. when 
$\lim_{t \rightarrow \infty} \la\dv_r^p(0)\ra\la\dv_R^q(t)\ra \equiv 0$
which is certainly false in
the most general case. The problem is not only 
limited to the necessity of taking
into account the asymptotic mismatch to zero given 
by the disconnected terms -- which would be a trivial modification
of (\ref{cf}) -- because as soon as the disconnected part
is present the whole hierarchy of fluctuating eddy-turn-over times
from the shortest, $t_R$, up to the largest,  $t_{L_0}$, must
be felt by the correlation. 

Let us, for the sake of simplicity, introduce a hierarchical set of 
scales, $l_n =2^n L_0$ with $n=0, \dots, n_d$, 
 which span the whole inertial range such that $l_d = l_{n_d} =2^{-n_d} L_0$,
and let us simplify the notation by taking $L_0=1$
and by  writing  $u_n =\dv_{r}$ in order
to refer to a velocity fluctuation at scale $r=l_n$.\\
The picture which 
will allow us to generalize the time-multifractal representation to the 
multi-time multi-scale case goes as follows.\\
For time-delays, $t \sim t_m $,
 typical of the eddy-turn-over time of the  $m$-th
scale  we may safely say that the two velocity fluctuations
follow the same fragmentation process from the
integral scale $L_0$ down to scale  $m$ while they
follow two uncorrelated processes from scale $m$ down to 
the smallest scale in the game, $n$. 
In the multifractal language we must write
that for time $t =t_m \pm O(t_m)$ 
we have:

\bea
u_n(0) \sim W'_{n,m}(0)W_{m,0}(0) u_0(0) \sim 
\left(\frac{l_n}{l_{m}}\right)^{h'} l_m^{h}\\
u_n(t) \sim W''_{n,m}(t)W_{m,0}(t) \sim W''_{n,m}(t)W_{m,0}(0)  u_0(0) 
\sim \left(\frac{l_n}{l_{m}}\right)^{h''}l_m^{h}
\eea
 where
the exponents $h,h',h''$ are independent outcomes 
of the same probability distribution functions and where we have used
the fact that in this time-window $W_{m,0}(t) \sim \mbox{const.}$\\
Apart from subtle further-time dependencies (see below) we should
therefore conclude that for time $t\sim t_m$ the correlation functions
may be approximated as:
\be
C_{n,n}^{p,q}(t_m) \sim \la W_{n,m}^p\ra\la W_{n,m}^q\ra\la W_{m,0}^{p+q}\ra
\label{kada}
\ee
which must be considered the fusion-rules prediction for the
time-dependent fragmentation process  
\cite{eyink,procaccia,bbt}. Let us notice that this 
proposal has already been presented  in \cite{kadanoff} 
and considered to express the leading term in the limit of large time
delays $t_m \rightarrow \infty$; here we want to refine the proposal
made in \cite{kadanoff} showing that by adding the proper time-dependencies
it is possible to obtain a coherent description of the correlation functions
for all time-delays. 
The expression (\ref{kada}) summarizes the idea that for time delay 
larger than $t_m$ but smaller then $t_{m-1}$, velocity components with
support on wavenumbers $k <k_{m-1}$ did not have enough time
to relax and therefore the local exponent, $h$, which describes  
fluctuations on those scales  must be the same for both fields. 
On the other hands, components with support on wavenumbers $k>k_{m-1}$ have 
already decorrelated for $t> t_{m-1}$ and therefore we must consider
two independent scaling exponents $h',h''$ for describing 
fluctuations on these scales.

Adding up all this fluctuations, centered  at different time-delays,
 we end with the following multifractal
representation for $C^{p,q}(l_n,l_n|t)\equiv C^{p,q}_{n,n}(t)$:
\begin{eqnarray}
C^{p,q}_{n,n}(t)& =& 
 \int d\mu_{m,0}(h)d\mu_{n,m}(h_1) d\mu_{n,m}(h_2) 
l_n^{(q+p)h} 
F_{p,q}\left(\frac{t}{t_n}\right)\,\,+
\label{mamma}\\
+\sum_{m=1}^{n-1} & \displaystyle{\int}& d\mu_{m,0}(h)d\mu_{n,m}(h_1)
d\mu_{n,m}(h_2) 
l_m^{(q+p)h} 
\left(\frac{l_n}{l_m}\right)^{qh_1}
\left(\frac{l_n}{l_m}\right)^{ph_2}
f_{p,q}\left(\frac{t}{t_m},\frac{t_m}{t_n}\right)
\nonumber  
\end{eqnarray}
Let us now spend a few words in order to motivate the previous expression.\\
In the first row of (\ref{mamma}) we have  explicitly
 separated 
the only  contribution
we would have in the case of vanishing disconnected part.
This term remains the leading contribution 
in  the static limit ($t=0$) also when disconnected parts are non-zero. 
About the new terms controlling the behavior of the correlation functions
for larger time we still have at this stage  the most general dependency
from all  times entering in the game $t,t_n,t_m$. 
In practice one may guess a very simple functional form which is 
in agreement with all the above mentioned phenomenological requirements.
In particular, we may simply assume that 
$f_{p,q}\left(\frac{t}{t_m},\frac{t_m}{t_n}\right) 
\equiv f_{p,q}\left(\frac{t}{t_m}\right)$ with $f_{p,q}(x)$ being a
function peaked at its argument  
$x \sim 0(1)$ which must be exactly zero for $x=0$
and different from zero only in  a  interval  of width $\delta x 
 \sim O(1)$.\\
Let us now face the consistency of the representation (\ref{mamma})
with the constraint imposed by the equations of motion.
By  applying 
a time-derivative to a correlation $C_{n,n}^{p,q}(t)$
you produce a new correlation with by-definition zero-disconnected part,
whose representation has thus no subleading term $(f_{p,q} \equiv 0)$.
When performing the time derivative on both sides of
(\ref{mamma}) it is evident that -- in order to accomplish the
dynamical constraints -- all the derivatives of the 
subleading terms must sum to a zero contribution. \\
This is the first non-trivial result we have reached until now.
If our representation (\ref{mamma}) is correct, we claim that 
all eddy-turn-over times must be  present in the general
single-scale correlation functions but 
strong cancellations of all sub-leading terms 
must take place whenever disconnected contributions  vanish.

Let us finally notice that for time
delays larger then the eddy-turn-over time of the integral scale
we should add to the RHS of (\ref{mamma}) the final exponential 
decay toward the full disconnected term $\la u_n^p\ra\la u_n^q\ra$.\\
In order to check this representation we have performed some numerical
investigation in a class of dynamical models of turbulence (shell models)
\cite{shell}. Without entering in the details of this popular
dynamical model for the  turbulent energy  cascade let us only say that
within this modeling the approximation of local-interactions among
velocity fluctuations at different scales is exact and therefore 
no-sweeping effects are presents. This fact makes of shell models
the ideal framework where non-trivial temporal properties 
can be investigated. \\
In order to test the dependency of  (\ref{mamma})
from the whole set of eddy-turn-over times  we show in Fig. 1
the correlation $C^{p,q}_{nn}(t)$ for two cases with and without
disconnected part. As it is evident, the correlation with a
non-zero disconnected part decays in a time-interval much longer
then the characteristic time of the shell $\tau_n$. 
This shows that it is not possible to associate a single time-scale
$\tau_n$ to the correlation functions of the form (\ref{mamma}). 
In Fig. 2 we also compare the correlation when one  
of the two observable is a time-derivative with the correlation
chosen such as to have the same dimensional properties but without
being an exact time-derivative. Also in this case the difference
is completely due to the absence (presence) of all sub-leading terms
in the former (latter). 

\subsection{Intermittent integral time-scales}
In the case when the disconnected part of the time correlation
is absent, in the representation (\ref{mamma}) all the subleading terms 
mutually cancel, leaving the fully-connected contribution alone.\\
Under this condition and in 
presence of intermittency one expects 
anomalous scaling behavior for 
the integral time-scales, $s^{p,q}(R)$, characterizing the mean decorrelation
time of fluctuations at scale $R$, defined as \cite{procaccia3}:
\be
s^{(p,q)}(R)= \frac{\int_0^{\infty} \; dt \; C^{p,q}(R,R|t)}{C^{p,q}(R,R|0)}
\label{ti}
\ee
exploiting the multifractal representation (\ref{mamma}) 
it is easy to show that:
\be
s^{(p,q)}(R) \simeq \left( \frac{R}{L_0} \right)^{z(p+q)}
\label{ti2}
\ee
where the exponents $z(m)$ are fully determined in terms
of the intermittent spatial scaling exponents: $z(m)=1+\zeta(m-1)-\zeta(m)$.\\
This prediction is in practice always
very difficult to check: indeed full
cancellation of the subleading terms requires an extremely long time span,
and since the cancellations affect dramatically the 
convergence of the time integral, there is no chance of measuring
with sufficient precision the $z(m)$ exponents.\\
In order to bypass this problem we 
devised an alternative way to extract the integral times.\\
We introduce fluctuating decorrelation times at a scale $R$, defined as 
the time interval $T_i^{(n)}$ in which the 
instantaneous value of the correlation has changed by a 
fixed factor $\lambda$, i.e. in our octave notation:
\be
u_n(t_i)u_n(t_i+T_i)=\lambda^{\pm 1}|u_n(t_i)|^2.  
\ee
At time $t_{i+1}=t_i+T_i$ we repeat this procedure
and we record the new decorrelation time $T_{i+1}$ and so forth
for an overall number of trials $N$.
The averaged decorrelation times can be thus defined as
$$\tau_n^{(m)}= 
\langle T^2 |u_n|^m \rangle_e / \langle T |u_n|^m \rangle_e
=\langle T |u_n|^m \rangle_t / \langle |u_n|^m \rangle_t, $$
where $\langle \cdots \rangle_e$ stands for ensemble averaging 
over the $N$ trials and $\langle \cdots \rangle$ represents 
the usual time average\footnote{The relation between the
e-average and the t-average is simply derived by observing that
 $\langle |u_n|^m T \rangle_t = 
(\int_0^{{\cal T}} T |u_n|^m dt)/{\cal T} \simeq
(\sum_i T_i^2 |u_n(t_i)|^m)/(\sum_i T_i)=
\langle T^2 |u_n|^m \rangle_e/\langle T \rangle_e$.}.
Since the multifractal description applies to time averages 
the averaged decorrelation times scale as $\tau_n^{(m)} \sim l_n^{z(m)}$ 
with the same scaling exponents of the integral times $s^{m}_n$. 

In Table 1  we report  the observed numerical $\zeta(m)$
along with the observed and expected scaling exponents for $\tau_n^{(m)}$,
showing a very good agreement. \\

\section{Two-scales time correlations}
\label{sec:4}
Let us now jump to the most general multi-scale multi-time correlation 
functions:
\begin{equation}
C_{N,n}^{p,q}(t) =\la u_n^q(0) \cdot u_N^p(t)\ra 
\label{msmt}
\end{equation}
where from now on we will always suppose that $u_N$ describes the velocity
fluctuation at the smallest of 
the two scales considered, i.e. $N>n$. 
 It is clear that now we have to consider the joint statistics
of two fields: the first, the slower,
at large  scale, $u_n(0)$ and the second, 
the faster, at small scale and at a time delay $t$, $u_N(t)$.

Following the same reasonings as before we may safely assume that 
from zero time delays up to time delays of the order of the slower velocity
field, $t=t_n$, the velocity field at small scale feels the same transfer
process of $u_n$ up to scale $n$ and then from scale $n$ to scale $N$
an uncorrelated transfer mechanism: 
\begin{equation}
u_N(t) =W_{N,n}(t) u_n(t)\sim W_{N,n}(t) u_n(0) \sim  W_{N,n}(t)W'_{n,0}(0)u_0 
\;
\;\; {\rm for } \; 0 \le t \le t_n
\label{fr}
\end{equation}
Similarly, for time delays within $t_n \le t_m < t < t_{m-1}\le t_0$
also the field at large scale $n$  will start to see different
transfer processes:
 \bea
u_n(0) \sim W''_{n,m}(0) u_{m}(0) \sim W''_{n,m}(0)W'_{m,0}(0)u_{0}\\
u_N(t) \sim W_{N,m}(t) u_m(t)\sim  W_{N,m}u_m(0) \sim
 W_{N,m}(t)W'_{m,0}(0)u_{0} 
\label{fr1}
\eea

It is clear now, how we may write down the correlation for any time:
\begin{eqnarray}
C^{p,q}_{N,n}(t) =&  & \label{mamma1}  \\
 \sum_{m=1}^n \int    d\mu_{m,0}(h)d\mu_{n,m}(h_1)
d\mu_{N,m}(h_2)&  
l_m^{(q+p)h} 
\left(\frac{l_n}{l_m}\right)^{qh_1}
\left(\frac{l_N}{l_m}\right)^{ph_2}&
f_{p,q}\left(\frac{t}{t_m},\frac{t_n}{t_m},\frac{t_N}{t_m}\right)
\nonumber 
\end{eqnarray}
where we want to stress that the sum in  the above expression goes
only up to the index of the largest scale $n$ (see below). 
In order to understand which would be a reasonable functional shape
for the $f_{p,q}$ function we need to point out  a few preliminary remarks.
Once we have to cope with two-scale correlation functions is natural
to suppose that the 
 the time-delay,
 $t_{nN}=t_n-t_N$,  needed for a energy burst to travel from shell
$n$ to shell $N$ will play an important r\^ole.
Furthermore, as far as the time-decaying properties are concerned
we must require that  only eddy-turn-over times
from the slower time $t_n$ up to the large-scale eddy-turn-over
time $t_0$ enter in the game. 
This is  because only for time larger then $t_n$ the correlation
is a true multi-time correlation. Indeed, for time-delay shorter then
$t_n$ only the field at small scale, $u_N$, 
 is changing but always under the same large scale
configuration,  $u_n$.\\ The final $f$-shape may be  therefore guessed as:
$f_{p,q}\left(\frac{t}{t_m},\frac{t_n}{t_m},\frac{t_N}{t_m}\right) = 
f_{p,q}\left(\frac{(t-t_{nN})}{t_m}\right)$. Where again the function $f(x)$
must be a function peaked for $x \sim 1$ and with a width $\delta x \sim O(1)$. 
The matching of representation (\ref{mamma1}) with the equation of
motion reveals  some important dynamical  properties. From simple
time-differentiation we should have  
\begin{equation}
\partial_t C_{N,n}^{p,q}(t) \sim O\left[\frac{C^{p+1,q}_{N,n}(t)}{l_N}\right]
\label{eq:dtfr} 
\end{equation}
which seems to be in {\it dis}agreement with the time-representation
proposed (\ref{mamma1}) because in the RHS of (\ref{eq:dtfr}) does appear
explicitly  the fast eddy-turn-over time $t_N$ (through the dependency
from $l_N$). Actually,
the representation (\ref{mamma1}) is still in agreement with the equation of
motion because the dependency of (\ref{eq:dtfr}) from $t_N$ is false:
again, exact cancellations must take place in the RHS. 
The explanation goes as follows:
in the multifractal language we may write $u_N(t) = W_{N,n}(t) u_n(t)$
and therefore
\be
\frac{d}{dt}
u_N(t) = \left(\frac{d}{dt}W_{N,n}(t)\right) u_n(t) + W_{N,n}(t)
\left(\frac{d}{dt}u_n(t)\right)
\label{dtfr2}
\ee but for
 time shorter then the eddy-turn-over, $t_n$, 
 of the  large scale $u_n$, the term 
$W_{N,n}(t) \left(\frac{d}{dt}\right)u_n(t)$  is zero because the shell $u_n$ 
did not move at all, while, once averaged,  the first term 
of the RHS of (\ref{dtfr2}) becomes 
$  \la\left(\frac{d}{dt}W_{N,n}(t)\right)\ra\la u_n(t)\ra$ which also vanishes
because of the total time derivative. The time derivative, $\partial_t
C_{n,N}^{p,q}(t)$ will therefore be a function which scales as
$\frac{C_{N,n}^{p,q+1}(t)}{l_n}$ instead of  
$\frac{C_{N,n}^{p+1,q}(t)}{l_N}$ as simple power counting would predict.
This may even be shown rigorously by evaluating the following averages:
\begin{equation}
\partial_t \la u_n^q(t)u_N^p(t)\ra \equiv 0 \equiv \la \left(\partial_t u_n^q\right)u_N^p\ra
+ \la u_n^q\left(\partial_t u_N^p\right)\ra.
\label{ward}
\end{equation}
The previous exact relation forces one of the two correlation
to not satisfy the simple multifractal ansatz because
otherwise  power-law counting
would be contradictory:
\begin{equation}
 \la\left(\partial_t u_n^q\right)u_N^p\ra \sim \frac{C_{N,n}^{p,q+1}(0)}{l_n} \neq
\la u_n^q\left(\partial_t u_N^p\right)\ra \sim \frac{C_{N,n}^{p+1,q}(0)}{l_N}
\end{equation}
Now, in view of the  previous discussion, we know that it is the correlation
with the time derivative at small scale, $\la u_n^q\left(\partial_t u_N^p\right)\ra$,
 which does not satisfy
the multifractal power law, but has the same scaling of 
$\la\left(\partial_t u_n^q\right)u_N^p\ra$ as our representation correctly reproduces. 

In order to test all these  properties, we plot in Fig. 3
the typical multi-time multiscale velocity correlation
$C_{N,n}^{p,q}(t)$ for $ p=q=1$, $n=6$, $N=6-13$. As one
can see the correlation has a peak which is in agreement with 
the delay predicted by (\ref{mamma1}), which saturates at the
value $\tau_{nN} \simeq \tau_{n}$ for $N \ll n$.
Let us also notice that due to the dynamical delay, $t_{m,N}$, the 
simultaneous multiscale correlation functions $C_{N,n}^{p,q}(0)$
do not show the fusion-rules prediction, i.e. pure power laws behaviors 
at all scales: 
\be
C_{N,n}^{p,q}(0) \sim \left(\frac{l_N}{l_n}\right)^{\zeta(p)}l_n^{\zeta(p+q)}
\label{frp}
\ee
Indeed, for $t \rightarrow 0$, the term $m=n$ dominates in (\ref{mamma1}) 
(because it makes $t_{nN}$ minimum) and
$f_{p,q}\left(-\frac{t_{nN}}{t_nm}\right)$
can be considered  a constant only in the limit of 
large  scale separation,  $n \ll N$, while 
otherwise we will see finite-size corrections.\\
The effect of the delay in multi-scale correlations is shown in 
Fig. 4  where we compare $C_{N,n}^{1,1}(0)$ and $C_{N,n}^{1,1}(T_{nN})$
(for $N>n=6$) rescaled with the Fusion Rule prediction (\ref{frp}). 
The time delay $T_{nN}$ is the time of the maximum of $C_{N,n}^{1,1}(t)$ 
computed from Fig. 3. We see that without delay, the prediction
(\ref{frp}) is recovered only for $N \gg n$ with a scaling factor 
$f_{1,1}(-1) \simeq 0.83$, while including the average delay $T_{nN}$ 
the Fusion Rule prediction is almost verified over all the inertial range.

Of course the delay $\tau_{nN}$ is a fluctuating quantity
and one should compute the average (\ref{mamma1}) with fluctuating
delays. In this case
the dimensional estimate $\tau_{nN}=l_n u_n^{-1}-l_N u_N^{-1}$
is somehow ill-defined, first of all being not positive definite. 
To find a correct definition for the fluctuating time delays is
a subtle point which lays beyond the scope of the present Paper.

\section{Conclusions}

In conclusion, we have proposed a multifractal-like
representation for the multi-time multi-scale velocity
correlation which should take into account all possible
subtle time-dependencies and scale-dependencies. The 
proposal can be seen as a merging of the proposal made 
in \cite{procaccia} -- valid only for cases when the disconnected
part is vanishing -- and the proposal made in \cite{kadanoff}
 -- valid only in the asymptotic regime of large
time delays and large scale separation. \\
Our proposal is phenomenologically
realistic and consistent with the dynamical constraints
imposed by the equation of motion. We have numerically 
tested our proposal
within the framework of shell models for turbulence. 

A new way to measure intermittent integral-time scales, $s^{p,q}(R)$,
has also been proposed and tested. \\
Further tests on the true Navier-Stokes eqs. would be of 
first-order importance. Furthermore, the building of
a  synthetic signals which would
reproduce the correct dynamical properties of the
energy cascade would also be of primary importance \cite{bbccv,bbct}.

We thank R. Benzi, V. L'vov and I. Procaccia for useful discussions.\\
This work has been partially supported by INFM 
(Progetto di Ricerca Avanzata TURBO). G.B. thanks the ``Istituto di Cosmogeofisica del CNR'', 
Torino, for hospitality.

\newpage
{\bf TABLE CAPTIONS}\\

TABLE 1: Comparison between the 
integral time-scales intermittency exponents, $z_m$, 
estimated from the measured spatial intermittent exponents, 
$z_m^{(th)}= 1 + \zeta(m-1)-\zeta(m)$, and from direct measuring via
the ``doubling-time'' $T$, $z_m^{(num)}$. \\

\newpage

{\bf FIGURE CAPTIONS}\\

FIGURE 1:\\
The 
time dependency of single-scale correlation functions, $C_{nn}(t)$,
in two different cases. The continuous line is  
the case with a non-zero disconnected part, $C_{nn}(t) = \langle 
|u_{n}(0)||u_{n}(t)|\rangle - \langle |u_{n}| \rangle^2$, 
while the dashed line represents a case with vanishing
disconnected part $\tilde{C}_{nn}(t)= \Re(\langle u_{n}(0)u^*_{n}(t) \rangle)$. 
Both correlations are rescaled to their value at $t=0$.
The scale is fixed in the middle of the inertial range, $n=12$,
 and the eddy turn over time of the reference scale was 
$\tau_{12} \simeq 0.29 $.  The average has been performed over approximately
 $500 \tau_{12}$, about $10$ large eddy turn over times.
The presence of subleading terms in $C_{nn}$ is apparent.
The remnant anticorrelation in $\tilde{C}_{nn}$, for $t>\tau_{12}$
reveals a partial cancellation of subleading terms: full cancellation
requires  averaging over a time interval of many more 
large eddy turn over times. 
\vspace{1 truecm}

FIGURE 2: \\
Comparison between the two correlations
$C_{nn}(t)=k_n \langle |u_n|^2(0) |u_n|^3(t) \rangle$, continuous line.
$D_{nn}(t)=-\langle |u_n|^2(0) \frac{d|u_n|^2}{dt}(t) \rangle$, dashed line.
The two correlations have the same dimensional properties, but 
$D_{nn}(t)$ decays faster due to cancellations of subleading terms.
$D_{nn}$ vanishes at zero delay because of stationarity and
smoothness of the process $u_n(t)$. 
Scale and characteristic times as in Figure 1.
\vspace{1 truecm}

FIGURE 3:\\
 Multi-time multi-scale correlation functions, $C_{n,N}(t)$,
for $n=6$ and $N=6, \cdots, 13$ (from bottom curve to top curve).
Observe the saturation in the time-delays, $\tau_{n,N} = \tau_n -\tau_N
\rightarrow \tau_n \simeq O(1)$
 when $N$ increases.\\

\vspace{1 truecm}
FIGURE 4: \\
Lin-log plot of 
Multi-scale correlation $C_{n,N}^{1,1}(t) = <|u_n(0)||u_N(t)|>$
 rescaled with  the Fusion Rule prediction: $C^{1,1}_nN(t)/ 
(S_N^1 S_{n}^{2}/S_{n}^{1})$ at fixed  $n=6$ and at changing 
$N \ge n$. The lower
line represent the zero-delay correlation ($t=0$), the upper line is for
the average delay $t=T_{6,N}$.

\newpage

\begin{table}
\begin{center}
\begin{tabular}{|c|c|c|}
\hline
$m$ & $\zeta_m$ &  $z_m^{(num)}$ $(z_m^{(th)})$ \\
\hline
 \hspace{3pt} 1 \hspace{3pt}   &  
 \hspace{3pt} 0.39 \hspace{3pt} & 
 \hspace{3pt} -0.61  (-0.61) \hspace{3pt} \\
2    &   0.72  &  -0.68  (-0.67) \\
3    &   1.00  &  -0.72  (-0.72) \\
4    &   1.26  &  -0.75  (-0.74) \\
5    &   1.49  &  -0.77  (-0.78) \\ 
6    &   1.71  &  -0.78  (-0.78) \\
7    &   1.93  &  -0.80  (-0.80) \\ 
8    &   2.13  &  -0.80  (-0.80) \\
\hline

\end{tabular}
\end{center}
\caption{}

\end{table}
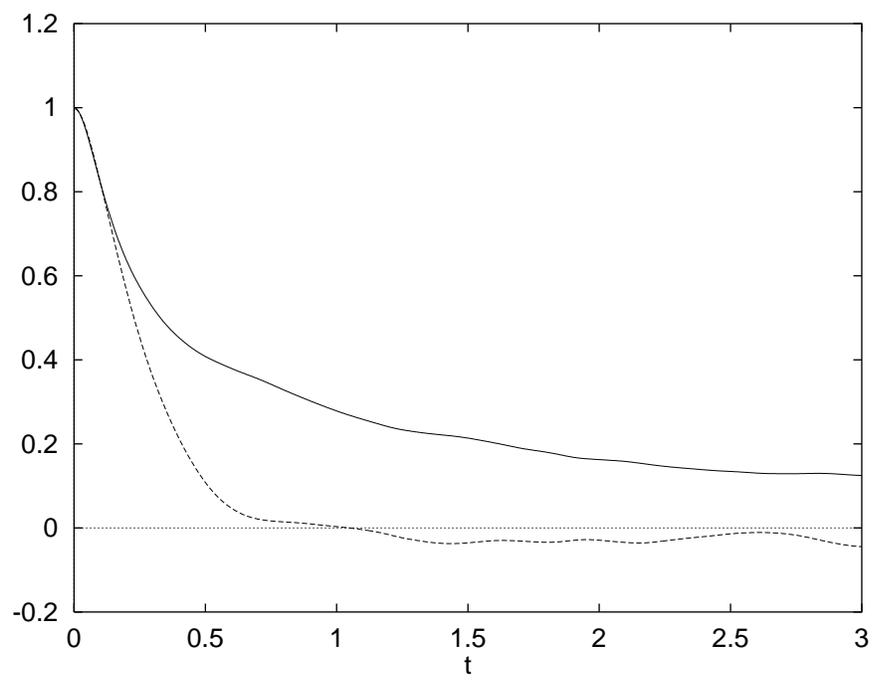
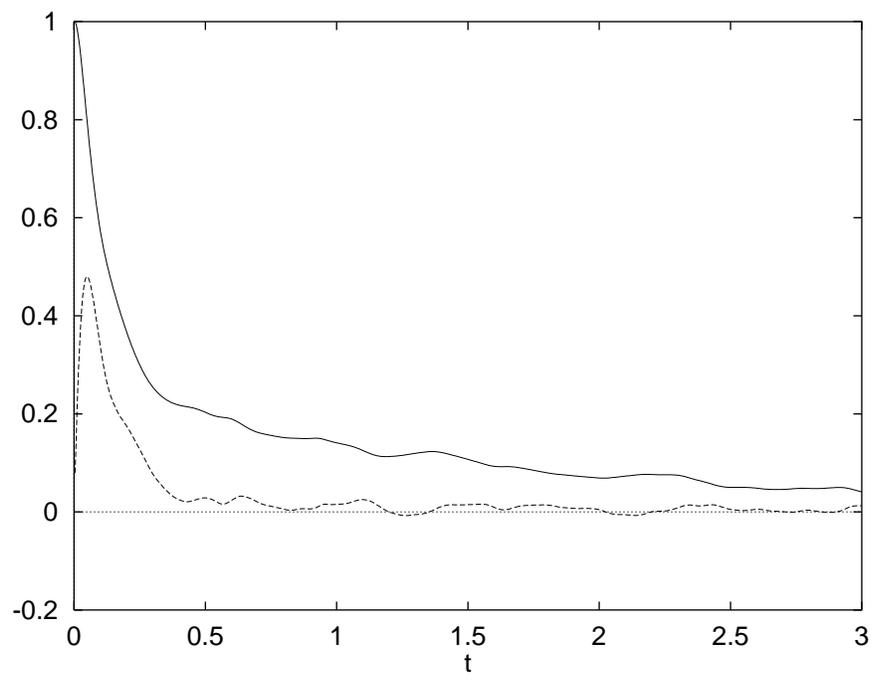
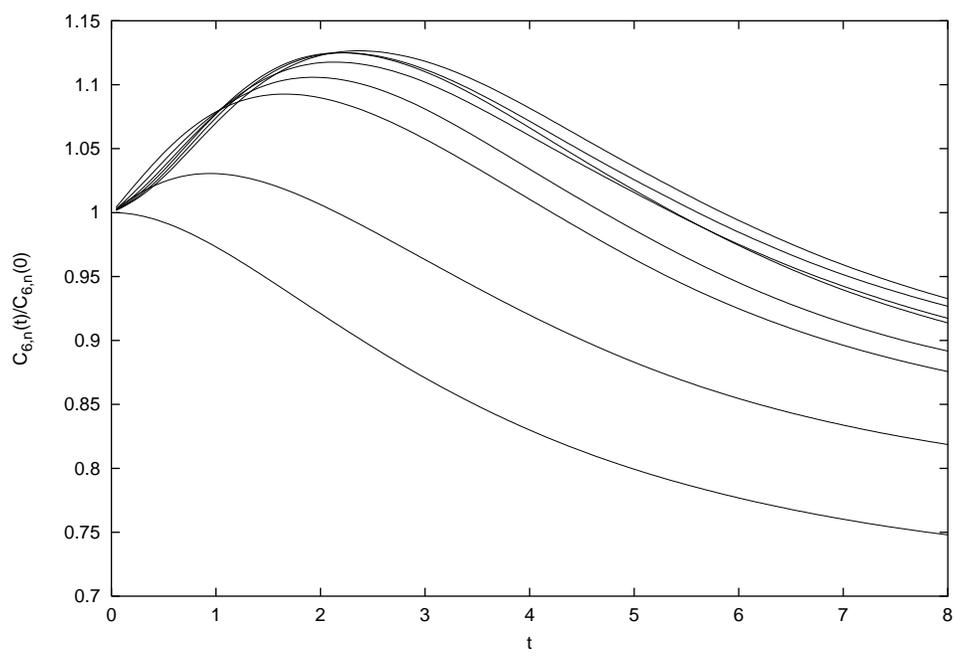
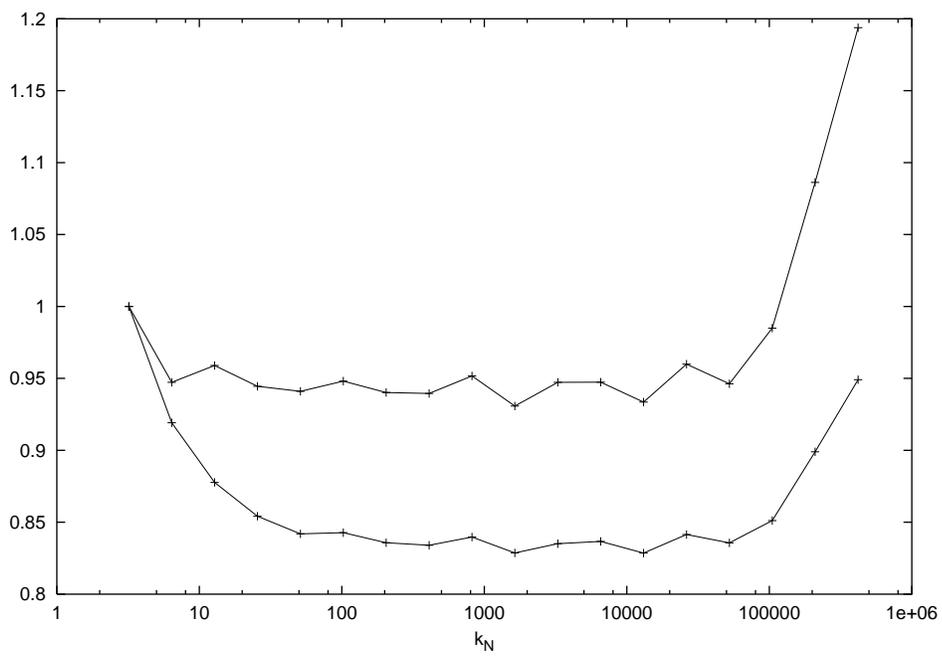
\end{document}